\documentclass[10pt,a4paper]{book}
\usepackage{frontier04}
 
\newcommand{\gray}{$\gamma$-ray}
\newcommand{\grays}{$\gamma$-rays}

\def\Xco{$X_{\rm CO}$}

\newcommand{\url}[1]{{\tt #1}}
\newcommand{\pubjournal}[6] {#1, #2 {\bf #3}, #4 (#5)}
\newcommand{\pubjournala}[6]{#1, #2, #4 (#5)}

\newcommand{\aap}{A\&A}
\newcommand{\adv}{Adv.Spa.Res.}
\newcommand{\apj}{ApJ}
\newcommand{\app}{Astropart.Phys.}
\newcommand{\apss}{Astrophys.Spa.Sci.}
\newcommand{\icrc}{ICRC}
\newcommand{\physrep}{Phys.Rep.}
\newcommand{\plb}{Phys.Lett.\ B}
\newcommand{\prd}{Phys.Rev.\ D}
\newcommand{\prl}{Phys.Rev.Lett.}
\newcommand{\ssr}{Spa.Sci.Rev.}

\begin{document}
\newcounter{ctr}
\setcounter{ctr}{\thepage}
\addtocounter{ctr}{10}

\talktitle{Propagation of Cosmic Rays and Diffuse Galactic Gamma Rays}
\talkauthors{Igor V.~Moskalenko} 

\begin{center}
\authorstucture[]{NASA Goddard Space Flight Center, Code 661,
                   Greenbelt, MD 20771, U.S.A.}

\authorstucture[]{Joint Center for Astrophysics/University of Maryland, Baltimore County 
                  \newline Baltimore, MD 21250, U.S.A.}
\end{center}

\shorttitle{Propagation of Cosmic Rays and Diffuse Gamma Rays} 

\firstauthor{Igor V.~Moskalenko}


\begin{abstract}
This paper presents an introduction to the astrophysics of cosmic rays 
and diffuse \grays\ and discusses some of the puzzles that have emerged 
recently due to more precise data and improved propagation models:
the excesses in Galactic diffuse \gray\ emission, secondary antiprotons and
positrons, and the flatter than expected gradient of cosmic rays in the Galaxy.
These also involve the dark matter, a challenge to
modern physics, through its indirect searches in cosmic rays.
Though the final solutions are yet to be found, I discuss some
ideas and results obtained mostly with the numerical propagation model GALPROP.
A fleet of spacecraft and balloon experiments targeting these specific 
issues is set to lift off in a few years, imparting a
feeling of optimism that a new era of exciting discoveries is just
around the corner.
A complete and comprehensive discussion of all the recent results
is not attempted here due to the space limitations.
\end{abstract}

\section{Introduction}

Research in astrophysics of cosmic rays (CR) and \grays\ 
provides a fertile ground for studies and discoveries in many areas of
particle physics and cosmology. Examples are the search for dark matter (DM),
antimatter, new particles, and exotic physics; studies of the
nucleosynthesis, acceleration of nuclei and their transport through
CR spectra and composition analysis; the effects of heliospheric modulation;
the origin of Galactic and extragalactic diffuse \gray\
emission; formation of the large scale structure of the
universe as traced by \grays.

In its turn, the astrophysics of CR and \grays\ depends very
much on the quality of the data and their proper interpretation.  The
accuracy of data from current CR experiments  on
interplanetary spacecraft such as Ulysses, Advanced Composition
Explorer (ACE), and the two Voyagers, specialized balloon-borne
experiments such as Super-TIGER, BESS-Polar, CREAM, and the new space-based
missions such as Pamela and AMS far exceeds the accuracy of the
current propagation models, of which the ``leaky-box" model has
remained one of the main research tools for the last 50 years. The new
major \gray\ observatory GLAST will improve the sensitivity for the
diffuse high-energy \grays\ produced in CR interactions in
the interstellar medium (ISM) by a factor of 30. These near-future missions are
specifically designed to search for the signatures of DM,
search for antimatter, study the diffuse Galactic and extragalactic
\grays, and provide outstanding quality data on \gray\ sources and
CR species in a wide energy range. This presents a great
opportunity for new discoveries that requires accurate, testable and
readily accessible modeling to exploit. 

On the other hand, the whole of our knowledge is
based on measurements done only at one point on the outskirts of the
Galaxy, the solar system, and the assumption that particle spectra and
composition are (almost) the same at every point of the Galaxy. The
latter may not necessarily be correct. \grays\ and
radio-waves (synchrotron) are able to deliver the
information directly from distant Galactic regions thus complementing that
obtained from CR measurements. Some part of the diffuse 
\grays\ is produced in energetic
nucleon interactions with gas via $\pi^0$ production,
another is produced by electrons via inverse Compton
scattering and bremsstrahlung. These processes are dominant in
different parts of the spectra of \grays, therefore, if deciphered the
\gray\ spectrum can provide information about the large-scale spectra of
nucleonic and leptonic components of CR. Combining
all the data collected by different experiments into a realistic
interpretive model of the Galaxy,
we have a better chance to understand the mechanisms of particle
acceleration and the role of energetic particles
in the dynamics and evolution of the Galaxy and make essential progress
in related areas.

\section{Cosmic Rays and Diffuse \grays}

CR are energetic particles which come to us from outer 
space and are measured through satellites, balloon-, and ground-based
instruments. The energy density of CR particles is about 1 eV cm$^{-3}$
and is comparable to the energy density of interstellar radiation
field (ISRF), magnetic field, and turbulent motions of the interstellar
gas. This makes CR one of the essential factors determining
the dynamics and processes in the ISM. 
The spectrum of CR can be approximately described by a
single power law with index --3 from 10 GeV to the highest energies
ever observed $\sim$$10^{21}$ eV. The only feature confirmed by observations of
many groups is a ``knee'' at $\sim$$3\times10^{15}$ eV. 
Meanwhile, the origin of the CR spectrum is not yet understood.

The sources of CR are believed to be supernovae (SNe) and 
supernova remnants (SNRs),
pulsars, compact objects in close binary systems, and stellar winds.
Observations of X-ray and \gray\ emission from these objects reveal
the presence of energetic particles thus testifying to efficient
acceleration processes in their neighborhood \cite{koyama-allen}.  
Particles accelerated near the sources
propagate tens of millions years in the ISM 
before escaping into the intergalactic space. In the course of
CR propagation,
secondary particles and \grays\ are produced, and the initial 
spectra of CR species and composition change. The destruction of
primary nuclei via spallation gives rise to secondary nuclei and
isotopes which are rare in nature, antiprotons, and pions
($\pi^\pm$, $\pi^0$) that decay producing secondary $e^\pm$'s and \grays.

Measurements of CR isotopic abundances are able to provide
detailed information about the acceleration mechanisms, source 
composition, and processes in the ISM. However, 
the energy range below 20 GeV/nucleon is strongly affected by the heliospheric
modulation, which is a combination of effects of convection by the solar
wind, diffusion, adiabatic cooling, different kinds of drifts, and
diffusive acceleration. The Pioneer, the two Voyagers, and Ulysses 
missions contributed significantly to understanding the
global aspects of modulation and limiting the number of modulation models'
free parameters, yet the relative importance of various
terms in the Parker equation is not established and appears 
to vary significantly over 22 year period \cite{potgieter}.
The most widely used are the
spherically symmetric force-field and Fisk 
approximations \cite{force-field-fisk}.

The diffuse \gray\
continuum emission is the dominant feature of the \gray\ sky. 
It is an evidence of energetic CR proton and electron
interactions with gas and the ISRF, 
and is created via $\pi^0$-decay, inverse Compton, and bremsstrahlung.
This emission in the range 50 keV
-- 50 GeV has been systematically studied in the experiments OSSE,
COMPTEL, and EGRET on the Compton Gamma-Ray Observatory as well as in
earlier experiments, SAS 2 and COS B \cite{hunter97-M04rev}. 
The observation of diffuse \grays\ provides the most direct test of the
proton and electron spectra on the large Galactic scale. 

\section{Models of Cosmic-Ray Propagation}

The analytical methods 
include the so-called leaky-box model and diffusion models
(e.g., disk-halo diffusion model, the dynamical halo wind model, the
turbulent diffusion model, reacceleration model). The leaky-box model
treats a galaxy as a box with reflecting boundaries and small leakage,
so that a particle travels across it many times before escaping. In this
model the principal parameter is an effective escape length or
grammage and particles, gas, and sources are distributed homogeneously.
The leaky-box model
has no predictive power or can even be wrong in cases when the distribution
of gas or/and radiation field is important, such as radioactive
isotopes (including K-capture isotopes), diffuse Galactic \grays,
electrons, positrons (because of their large energy losses) etc.
It also can not account for spatial variations of CR intensity. 
Diffusion models \cite{jones01a} are more realistic, distinguishing
between the thin Galactic disk and extensive halo, often with
different diffusion  coefficient. The technique used, e.g., the weighted
slab technique, which splits the problem into astrophysical
and nuclear parts, may, however, give significant errors in some cases.
The alternative way is the direct numerical solution of the diffusion
transport equations for the entire Galaxy and for all CR
species. 

The modeling of CR diffusion in the Galaxy includes the
solution of the transport equation with a given source distribution
and boundary conditions for all CR species. The transport
equation describes diffusion and energy losses and may also include
\cite{Zirakashvili96-seo-ptuskin03} the
convection by a hypothetical Galactic wind,
distributed acceleration in the ISM due to the Fermi second-order mechanism, 
and non-linear wave-particle interactions.
The boundary conditions assume free particle escape into the intergalactic
space.

The study of stable secondary nuclei (Li, Be, B, Sc, Ti, V) 
allows one to determine the ratio (halo size)/(diffusion
coefficient) and the incorporation of radioactive secondaries 
($^{10}_{4}$Be, $^{26}_{13}$Al, $^{36}_{17}$Cl, $^{54}_{25}$Mn) is used to
find the diffusion coefficient and the halo size \cite{ptuskin-webber,SM98}.
The derived source abundances of CR
may provide some clues to mechanisms and sites of CR acceleration.
However, the interpretation of CR data, e.g.,
the sharp peak in the secondary/primary nuclei 
ratio (e.g., B/C), depends on the adapted physical model.
The leaky-box model fits the secondary/primary ratio by 
allowing the path-length distribution vs.\ rigidity to vary.
The diffusion models are more physical and explain the shape of
the secondary/primary ratio in terms of
diffusive reacceleration in the ISM, 
convection by the Galactic wind,
or by the damping of the interstellar turbulence by CR on a small scale.

K-capture isotopes in CR 
(e.g., $^{49}_{23}$V, $^{51}_{24}$Cr) can 
serve as important energy markers and can be used to study
the energy-dependent effects. Such nuclei usually 
decay via electron-capture and have a short lifetime
in the medium. In CR they are stable or live
essentially longer as they are created bare by fragmentation
of heavier nuclei while their $\beta^+$-decay mode
is suppressed.
At low energies, their lifetime depends on the balance between
the processes of the electron attachment from the interstellar gas
and stripping thus making
their abundances in CR energy-dependent.
This opens a possibility to probe the diffusive 
reacceleration in the ISM and heliospheric
modulation \cite{soutoul98-niebur03-jones01b}.

The study of transport of the CR nuclear component
requires the consideration of nuclear spallation,
radioactive decay, and ionization energy
losses. Calculation of isotopic abundances involves
hundreds of stable and radioactive isotopes produced in
the course of CR interactions with interstellar gas.
A thorough data base of isotopic
production and fragmentation cross sections and particle data is thus
a critical element of models of particle propagation
that are constrained by the abundance measurements of isotopes, antiprotons,
and positrons in CR. 
Meanwhile, the accuracy of many of the nuclear cross sections used in
astrophysics is far behind the accuracy of the current CR experiments, 
such as Ulysses, ACE, and
Voyager, and clearly becomes a factor restricting further progress.
The widely used semi-phenomenological systematics have typical
uncertainties more than $\sim$30\%, and can sometimes be wrong by an
order of magnitude \cite{spallation-yanasak-MMS01-MM03};
this is reflected in the value of propagation parameters. 

Increasingly accurate balloon-borne and
spacecraft experiments justify the development of sophisticated and
detailed propagation models with improved predictive capability. 
Ideally, such a model has to incorporate all
recent developments in astrophysics, such as detailed 3-dimensional maps of
the Galactic gas derived from radio and IR surveys, the
Local Bubble structure and local SNRs, the spectrum and intensity of the ISRF,
Galactic magnetic fields, details of composition of interstellar dust, grains, 
as well as theoretical work on particle acceleration and transport in Galactic 
environments. A detailed gas distribution is important for accurate calculations 
of the spectra of $e^\pm$'s, radioactive species, 
and for calculation of \gray\ flux and skymaps 
from electron bremsstrahlung and from the decay of $\pi^0$'s produced by CR
interactions. The ISRF is essential for electron and positron
propagation (energy losses) and \gray\ production by inverse Compton
scattering. The magnetic field provides
useful constraints on the electron spectrum via synchrotron emission,
and may establish preferential directions of propagation of CR particles.
Inclusion of the Local Bubble and SNRs enables us to study CR 
intensity and spectral variations in the local ISM.

A well-developed and sophisticated propagation model,
in return, provides a basis for many studies in
astrophysics, particle physics, and cosmology. 
The indirect DM search is a good example. 
A clear feature found in the spectra of CR antiprotons, 
positrons, or diffuse \grays\ would be a ``smoking gun'' 
for DM \cite{bergstrom01_prl}; but nature is unlikely to be so cooperative.
A more reasonable expectation is that the DM signature, if any at all, 
will be a weak broad signal on top of a background requiring a
reliable propagation model to be able to discriminate the signal.

Modern computer codes incorporating recent developments in
astrophysics and nuclear physics do exist.
One example is GALPROP \cite{SM98,SMR00,M02}, a numerical model and
a computer code (written in C++) of CR propagation in the Galaxy; 
this is the most advanced 3-dimensional model to date. 
The model is designed to perform CR propagation calculations for nuclei 
($_1^1$H to $_{28}^{64}$Ni), antiprotons, electrons, positrons, and 
computes \grays\ and synchrotron emission in the same framework; 
it includes all relevant processes and reactions.
The GALPROP model has been proven to be a useful and powerful tool to
study the CR propagation and related phenomena. Its
results (and the code) are widely used as a basis for many studies, 
such as search for DM
signatures, origin of the elements, the spectrum and origin of Galactic
and extragalactic diffuse \gray\ emission, heliospheric modulation
etc. The GALPROP code, or components of it, are being used by the
members of experimental teams, such as GLAST, AMS, Pamela,
HEAT, ACE, TIGER, and requested by many other researchers world-wide.

\section{Science Frontiers in Astrophysics of Cosmic Rays}

\subsection{Diffuse \grays\ and Cosmic-Ray Gradient}
The puzzling excess in the EGRET data above 1 GeV relative to that
expected \cite{hunter97-M04rev} has shown up in all models that are tuned to
be consistent with local nucleon and electron spectra
\cite{SMR00,SMR04a-SMR04b}. Is it a key to the problems of CR physics,
an evidence of the Local Bubble,
a signature of exotic physics (e.g., WIMP annihilation, primordial
black hole evaporation), or just a flaw in the current models?  This
also has an immediate impact on the extragalactic background radiation
studies since its spectrum and interpretation are model dependent.
An apparent discrepancy between the radial
gradient in the diffuse Galactic \gray\ emissivity and the
distribution of CR sources (SNRs) has worsened the problem 
\cite{SMR00}.

The puzzle of the ``GeV excess'' has lead to an attempt to
re-evaluate the reaction of $\pi^0$-production in
$pp$-interactions. 
A calculation made using Monte Carlo event generators to simulate high-energy
$pp$-collisions confirmed previous results \cite{mori97}. 
A parametrization \cite{blattnig} gives larger
number of $\pi^0$'s produced at high energies
compared to a standard formalism \cite{StephensBadhwar81}
while consistent with pion data; 
its effect on diffuse \grays\ is not studied yet.

Another
leading reason for the discrepancy discussed is that the local CR
particle spectra (nucleons and/or electrons) may be not
representative of the Galactic average.  The local source(s) and
propagation effects (e.g., energy losses) can change the
spectrum of accelerated particles.
A flatter Galactic nucleon spectrum has been suggested as a possible
solution to the ``GeV excess'' problem
\cite{mori97,gralewicz97}. This requires the
power-law index of proton spectrum of about --2.4--2.5, however, it
is inconsistent with measurements of 
CR $\bar p$ and $e^+$ fluxes \cite{MSR98}.
Besides, the GeV excess appears in all directions
\cite{SMR04a-SMR04b} implying that this is not a feature
restricted to the gas-related emission. A flatter (hard)
electron spectrum, justified by the random nature of CR
sources and large energy losses, may explain the GeV excess in terms of inverse 
Compton emission \cite{porter97-pohl98-aharonian}. However,
the required fluctuations are too large and the calculated
spectrum of diffuse \grays\ is inconsistent with EGRET data 
above 10 GeV \cite{SMR04a-SMR04b}.

In the new analysis of the Galactic diffuse \gray\ emission \cite{SMR04a-SMR04b}, 
CR $\bar p$ data were used to fix the Galactic
average proton spectrum, while the electron spectrum is adjusted using
the spectrum of diffuse \grays\ themselves. The derived electron and
proton spectra are found to be \emph{compatible} with those measured 
locally considering fluctuations due to energy losses, propagation, 
or possibly details of Galactic structure.
The effect of anisotropic inverse Compton scattering in the halo
can increase the high-latitude Galactic \gray\ flux 
up to 40\% \cite{MS00a}.
The model shows a good agreement with EGRET spectra of diffuse \gray\
emission from different sky regions ($<$100 GeV).
Some part of the excess can be associated
with SNRs where freshly accelerated particles
strike gas particles nearby producing harder \gray\ spectra \cite{Berezhko04}.
The increased Galactic contribution to the diffuse emission reduces
an estimate of the extragalactic \gray\ background \cite{SMR04a-SMR04b}.
The new extragalactic background shows a positive
curvature, which is expected if the sources are unresolved blazars
or annihilations of the neutralino DM \cite{salamon-mannheim}. 

The discrepancy between the radial
gradient in the diffuse Galactic \gray\ emissivity and the
distribution of SNRs \cite{case98-lorimer04}, believed to be the CR sources,
can be plausibly solved \cite{SMR04c}
if the \Xco-factor ($\equiv N_{{\rm H}_2}/W_{\rm CO}$) increases by a 
factor of 5--10 from the inner to the
outer Galaxy. The latter is expected from the Galactic metallicity 
gradient. 

\subsection{Secondary Antiprotons in CR}
Secondary antiprotons are produced in the same interactions 
of CR particles with interstellar gas as positrons and diffuse \grays. Their
unique spectral shape is seen as a key link between physics of CR
and diffuse \grays\ and could provide important clues to such
problems as Galactic CR propagation, possible imprints of our
local environment, heliospheric modulation, DM etc.

New $\bar p$ data with larger statistics
\cite{Orito00-Maeno01-beach-boezio01-bergstrom} 
triggered a series of calculations of the secondary $\bar p$ 
flux in CR.
The diffusive reacceleration models have certain 
advantages compared to other propagation models: they naturally reproduce
secondary/primary nuclei ratios in CR, have only three
free parameters (normalization and index of the diffusion
coefficient, and the Alfv\`en speed), and agree better with
K-capture parent/daughter nuclei ratio.
The detailed analysis shows, however, that the reacceleration models 
produce too few $\bar p$'s \cite{M02} because
matching the B/C ratio at all energies requires the diffusion
coefficient to be too large. 
The discrepancy in $\bar p$ flux is $\sim$40\% at 2 GeV.

The difficulty associated with antiprotons may indicate new effects. 
It may indicate \cite{M02} that propagation
of low-energy particles is aligned to the magnetic field lines 
instead of isotropic diffusion. If our local environment (the Local 
Bubble) influences the spectrum of CR, then the problem can be solved by
invoking a fresh ``unprocessed'' nuclei component
at low energy \cite{M03-M03icrc}; the evidence for SN
activity in the solar vicinity in the last few Myr supports this idea.
More intensive CR flux in distant regions will also 
produce more antiprotons and diffuse \grays\ \cite{SMR04a-SMR04b}. 

The computed interstellar flux of secondary antiprotons 
can be used to test the models of solar modulation.
Using a steady-state 
drift model of propagation in the heliosphere,
the predictions are made for $p$'s and $\bar p$'s fluxes near the
Earth for the whole 22 year solar cycle \cite{M02,potgieter04a-potgieter04b}; 
this includes different modulation levels and 
magnetic field polarities.

\subsection{Indirect Searches for Dark Matter}

The nature of the non-baryonic DM is a mystery. One of the
preferred candidates for non-baryonic DM is a weakly interacting
massive particle (WIMP).
In most models the WIMP is the lightest neutralino $\chi^0$
\cite{jkg-bergstrom00}, which arises naturally in supersymmetric
extensions of the Standard Model of particle physics. 
Another candidate is a Kaluza-Klein particle \cite{matchev}, a hypercharge $B^1$
gauge boson, whose thermal relic density is consistent with the WMAP
measurements. Annihilations of DM particles produce
leptons, quarks, gluons, and bosons, which 
eventually decay to ordinary particles.
The DM particles in the Galactic halo 
or at the Galactic center \cite{gunn-stecker-silk-gondolo-silk-gondolo} may thus be
detectable via their annihilation products ($e^+$, $\bar p$, $\bar d$, 
\grays) in CR 
\cite{bergstrom99-baltz99-baltz03-darksusy04}. The approach is to scan the
SUSY parameter space to find a suitable candidate particle to fill the 
excesses in diffuse \grays, $\bar p$'s, and
$e^+$'s over the predictions of a conventional model. 
A preliminary results of the ``global fit'' to the 
$e^+$'s, $\bar p$'s, and diffuse \gray\ data simultaneously look promising 
\cite{deboer1-deboer2}.

\section{Conclusion}

The choice of topics
discussed in this paper is personal and by no means complete.
More complete list would include the origin of 511 keV line from the
inner Galaxy, \gray\ bursts, ultra-high energy
CR, as well as a more comprehensive discussion of
the DM, SUSY, and dark energy. Other contributions 
to the Conference will fill these gaps.

This work was supported in
part by a NASA Astrophysics Theory Program grant.

\end{document}